\documentclass[aps, prl, twocolumn, floatfix]{revtex4-1}
\usepackage{amssymb, amsmath, bm, graphicx, color, colortbl, xfrac, tabularx}
\usepackage{todonotes}
\DeclareGraphicsExtensions{.pdf,.png,.jpg, .eps}

\definecolor{Gray}{gray}{0.9}
\definecolor{LGray}{gray}{0.8}

\begin{document}
\title{Braiding and fusion of non-Abelian vortex anyons}
\author{T. Mawson*$^{1}$}
\author{T. C. Petersen$^{1}$}
\author{J.~K. Slingerland$^{2,3}$}
\author{T.~P. Simula$^{1,4}$}
\affiliation{$^{1}$School of Physics and Astronomy, Monash University, Victoria 3800, Australia}
\affiliation{$^{2}$Department of Theoretical Physics, Maynooth University, Co. Kildare, Ireland}
\affiliation{$^{3}$Dublin Institute for Advanced Studies, School of Theoretical Physics, 10 Burlington Rd, Dublin, Ireland}
\affiliation{$^{4}$Centre for Quantum and Optical Science, Swinburne University of Technology, Melbourne 3122, Australia}

\begin{abstract}
We demonstrate that certain vortices in spinor Bose-Einstein condensates are non-Abelian anyons and may be useful for topological quantum computation. We perform numerical experiments of controllable braiding and fusion of such vortices, implementing the actions required for manipulating topological qubits. Our results suggest that a new platform for topological quantum information processing could potentially be developed by harnessing non-Abelian vortex anyons in spinor Bose–-Einstein condensates.
\end{abstract}

\maketitle

All elementary particles are classified by their quantum statistics as either bosons or fermions. However, in certain two-dimensional materials, particle-like excitations---anyons---which are neither bosons or fermions, have been predicted to emerge \cite{Leinaas1977,PhysRevLett.48.1144}. When two anyons are exchanged, braiding their space-time world-lines, the system's wave function may accumulate an arbitrary phase not restricted to the specific values corresponding to bosons or fermions. For non-Abelian anyons, exchange may act through non-commuting unitary operators, rather than simple phases. Also, how such anyons fuse (combine) when brought together depends on the history of their paths prior to the fusion.  Encoding information in the non-local fusion properties of non-Abelian anyons forms a tantalising prospect for realisation of a fault-tolerant universal quantum computer \cite{Freedman1998a,Kitaev2003a}.  

Recent advances in quantum computing have come from intense research focus on qubits realised in a variety of systems including trapped ions \cite{CiracZoller1995a,ospelkaus_microwave_2011,Nigg302}, spins in silicon atoms \cite{RevModPhys.85.961,Watson2018a} and superconducting circuits \cite{clarke_superconducting_2008,Wendin2017a}. Such systems must contend with the accumulation of spontaneous errors due to the inherent interactions of the qubits with their environment. In contrast, topological quantum computers based on topological qubits made of non-Abelian anyons are anticipated to be far more resilient due to being topologically protected from many conventional types of decoherence. Two promising non-Abelian anyon platforms are the Fibonacci and Ising anyon models \cite{KITAEV20062,trebst_short_2008, RevModPhys.80.1083,Sarma2015a,Field2018a}. A number of experiments have explored the potential realisation of such anyons in condensed matter systems including Majorana zero modes \cite{Mourik1003,Nadj-Perge602,He294,zhang_quantized_2018,Zhang2018a,Zhang2018a} and quasiparticles in certain fractional quantum Hall states \cite{Radu899,cite-key,Willett8853}. Other non-Abelian anyon models have been proposed to be realisable using \emph{fluxons} \cite{Bais1980a,BREKKE1993127,PhysRevD.48.4821}. Notwithstanding, the existence of a physical system of non-Abelian anyons capable of universal quantum computation remains an open question.

Here we computationally demonstrate that certain fractional vortices---particle-like topological excitations in two-dimensional (2D) spinor Bose--Einstein condensates (BECs)---may be non-Abelian anyons and are potentially useful for applications in topological quantum information processing and storage. In addition to fluxons, excitations in these systems include \emph{chargeons} \cite{preskill} and charge-flux composites known as \emph{dyons} \cite{Schwinger1966a,Schwinger1968a,Zwanziger1968a,Witten1979a}. The full spectrum of excitations is labeled using the quantum double of the symmetry group of the condensate
\cite{deWildPropitius, *deWildPropitius1995a}. In addition to chargeons, these systems also allow for completely delocalized \emph{Cheshire charges} \cite{Alford1990a,Preskill1990a}. 
We simulate the braiding and fusion of non-Abelian vortex anyons by employing external pinning potentials that could be realised using focused laser beams \cite{Roberts2014a,Samson2016a}, to controllably manipulate the states of topological qubits constructed from such non-Abelian vortex anyons. 

\begin{figure*}[ht!]
\centering
\includegraphics[width=1.5\columnwidth]{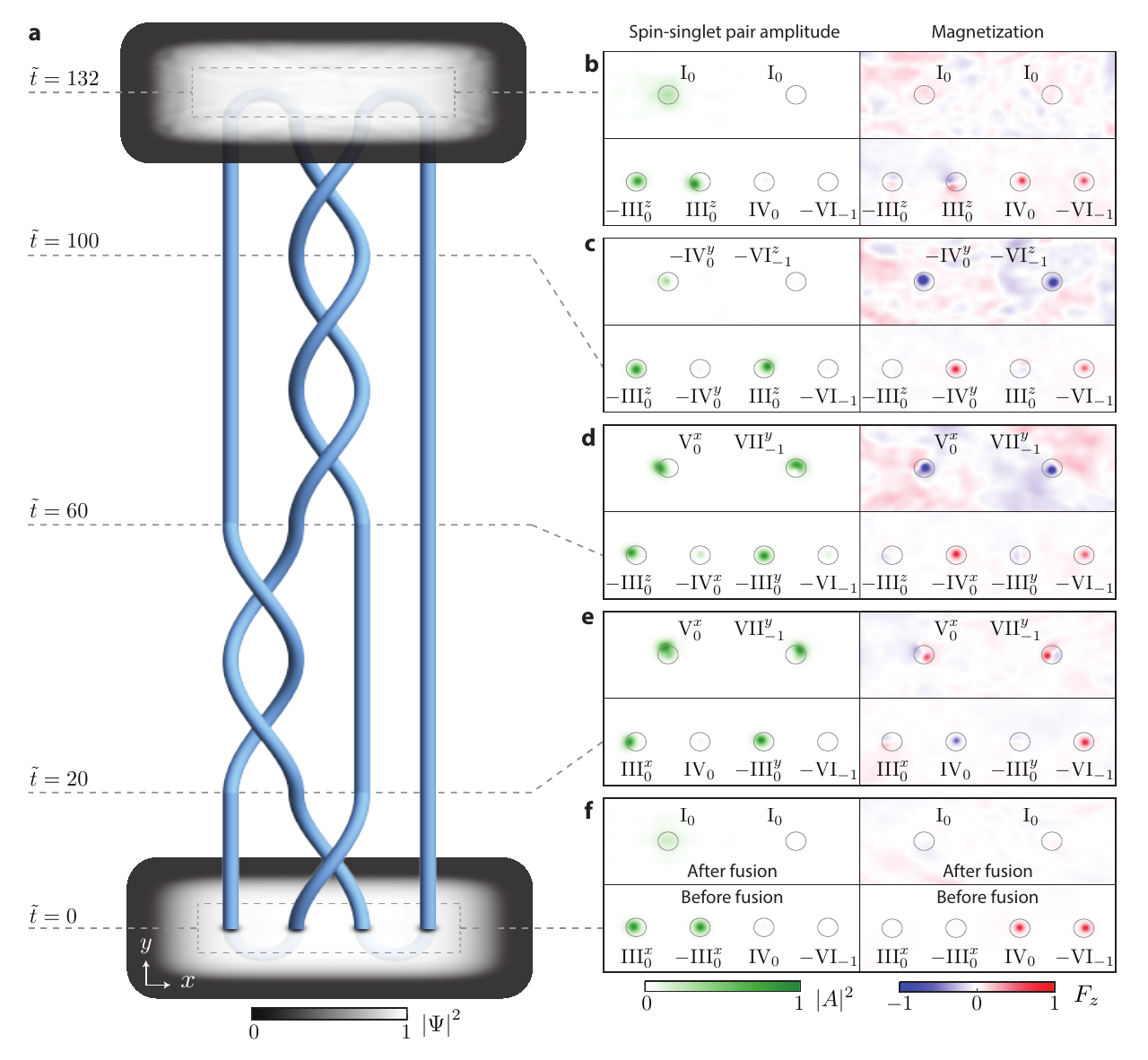}
\caption{\textbf{Braiding and fusion of non-Abelian fractional vortices}. \textbf{a}, The paths of vortices embedded in a two-dimensional Bose--Einstein condensate trace out world lines that form a braid whose plat closure yields a link (Supplementary Video 1). The total condensate density is shown for the initial ($\tilde{t}=0$) and final ($\tilde{t}=132$) states. \textbf{b}, Spin-singlet pair amplitude (left column) and magnetization (right column) with vortex locations marked using circles and labelled by the vortex (anyon) types. The upper rows correspond to the system state just after the vortices have been fused pairwise and the lower rows correspond to the state just before the fusion. 
The field of view of each of the four frames in {\bf b}-{\bf f} corresponds to the dashed rectangle shown in {\bf a} where the inter-vortex separation is 27$\mu$m. The dimensionless times $\tilde{t} =t \omega$ of measurement of states {\bf b}-{\bf f} are  marked in {\bf a}.
}
\label{L6a2}
\end{figure*}

\textit{Non-Abelian vortex anyons---}A non-Abelian anyon model has three essential aspects; (i) a list of particle types; (ii) a set of fusion rules that determine the types of particles formed after fusing together two particles; and (iii) braiding rules that describe the effect of exchanging the positions of two particles. We demonstrate that the topological interactions of our non-Abelian fractional vortices in spinor Bose--Einstein condensates \cite{Ciobanu2000a,0305-4470-39-23-017,semenoff_discrete_2007,PhysRevLett.103.115301,Huhtamaki2009a,Kawaguchi2012253,PhysRevLett.117.275302,PhysRevA.96.033623} contain the essential aspects of a non-Abelian anyon model. The anyon models involved are similar to those of non-Abelian toric code models \cite{Kitaev2003a} or discrete gauge theories \cite{deWildPropitius, *deWildPropitius1995a}. Each non-Abelian vortex is characterized by a distinct topological flux represented by a matrix that corresponds to a certain transformation of the spinor wave function of the Bose--Einstein condensate, as it winds around the core of the vortex. The transformation matrices that represent fluxes form a group, the stabiliser group $H$ of the condensate. A specific flux can be represented by a particular transformation or by any of its conjugates within this group. Distinguishable flux types correspond to fluxons and are labelled by the sets of conjugate transformations---the conjugacy classes of $H$. The fluxons represent distinguishable topological particle types in the theory. In addition to fluxons, there are $H$-charges or chargeons, labelled by representations of $H$. Physically, the fluxon particles in our system are associated with vortices, and more specifically the low-energy Bogoliubov quasiparticle eigenstates localized within a vortex core \cite{Simula2008a}. Similarly, the chargeon particles may be viewed as Bogoliubov `zero' modes, such as those associated with a phonon, a magnon or a soliton, which are organized into multiplets under the action of $H$ (which leaves the state of the condensate invariant).

Physically, vortices with non-commuting topological fluxes are characterised by non-trivial, path dependent, topological interactions. Figure~\ref{L6a2} shows the outcome of a numerical experiment, obtained by solving the spin-2 Gross--Pitaevskii equation (see Methods) in 2D governed by a five component spinor wavefunction $\Psi$, that demonstrates the exotic braiding and fusion dynamics of non-Abelian vortices. The system is initialized at time $t=0$ in Fig.~\ref{L6a2}a by creating four non-Abelian vortices in the Bose--Einstein condensate by phase-imprinting two vortex-antivortex pairs, one on the left and one on the right hand side of the rectangular condensate. Using pinning potentials that model an array of Gaussian--shaped laser beams that repel atoms, the vortices can be pinned and controllably moved around, forming a braid in their space-time world lines as shown in Fig.~\ref{L6a2}a and in Supplementary Videos. A plat closure of the braid is realised by the initial pair-creation and final fusion of the vortex pairs. The effects of braiding the vortices are observed at different dimensionless times $\tilde{t} = t\omega$, where $\omega= 2\pi\times5$ Hz, after alternatively, (i) releasing the pinning potentials and measuring the properties of the four vortices, see lower rows in Fig.~\ref{L6a2}(b-f), or, (ii) fusing the two vortex pairs first and then measuring the result after releasing the pinning potentials, see upper rows in Fig.~\ref{L6a2}(b-f). The vortex locations are visualised via their core structure, which may have non-zero spin-singlet pair amplitude, $|A|^2$, and/or non-zero magnetisation, $F_z$ (see Methods). 

A detailed understanding of the observed dynamics comes from labelling the flux of each vortex in Fig.~\ref{L6a2}(b-f), enabled by the vortex identification method described in Ref.~\cite{PhysRevA.96.033623}, which is based on the spherical harmonics representation of the spinor order parameter. The vortex labels are elements of the group $H$. Here, they have the form $\pm\rm{X^{\nu}_\eta}$, where $\rm{X}_\eta$ is a Roman numeral denoting the fluxon part of the anyon type. Each anyon may represent several vortices whose topological fluxes are further specified by the combination of symbols $\nu$ and $\eta$, and the $\pm$ sign (see Methods). Underpinning the braiding dynamics is the long-range topological influence between non-Abelian vortices  \cite{BREKKE1993127,PhysRevD.48.4821,kobayashi_topological_2014}. For an anti-clockwise elementary braid (exchange of a pair) of vortices with fluxes $\left(\gamma_1,\gamma_2\right)$ their mutual topological influence converts their fluxes to $(\gamma_2, \gamma_2\gamma_1\gamma_2)$,
see Supplementary Figure~S1. The products and the inverse in $\gamma_2\gamma_1\gamma_2^{-1}$ are taken in the group $H$.  If $\gamma_1$ and $\gamma_2$ do not commute, this mapping permutes the flux of the second vortex within the set of fluxes associated with its anyon type. The clockwise exchange realises the map $(\gamma_1,\,\gamma_2) \rightarrow (\gamma_1^{-1}\gamma_2\gamma_1,\,\gamma_1)$. Braiding may also enact a local unitary transformation on the wave function, which reverses the sign of the vortex core magnetisation, turning a red core into a blue core, and vice versa, without changing the value of their fluxes, as shown in Fig.~\ref{L6a2}e and ~\ref{L6a2}d. The outcome of fusing two vortices is determined by an ordered product of their two fluxes equivalent to their total flux. Only vortices whose fluxes multiply to the identity element of $H$ may annihilate, otherwise the fusion results in a remnant vortex. It is also possible for vortices with commuting fluxes to pass through each other without apparent interaction \cite{Bais1980a,PhysRevA.96.033623}.  

The initial vortex-antivortex pairs in Fig.~\ref{L6a2}f (lower row) consist of three particle types; two vortices of same type (III$_0$) with non-zero $|A|^2$, green cores, and two of different types (IV$_0$ and VI$_{-1}$) with $F_z > 0$, red cores. Initially, both pairs annihilate upon being fused (Fig.~\ref{L6a2}f, upper row), by construction. An exchange of the two vortices in the middle leads to the state measured at $\tilde{t} = 20$, shown in Fig.~\ref{L6a2}e. The braid swaps the positions of two vortices, which trivially changes the pairwise fusion dynamics as neither the green and red or green and blue cored vortices can annihilate. The braid between $\tilde{t}=60$ and $\tilde{t}=100$ consists of two exchanges (elementary braids) of the two middle vortices resulting in the state shown in Fig.~\ref{L6a2}c. Importantly, although this braiding preserves the ordering of the vortex types by returning them to their original pre-braiding positions at $\tilde{t}=60$, the types of vortices formed after fusion are different before (V$_0$ and VII$_{-1}$ at $\tilde{t}=60$) and after (IV$_0$ and VI$_{-1}$ at $\tilde{t}=100$) the braiding. Such vortex metamorphosis due to braiding is a hallmark of non-Abelian anyons. The final exchange of the middle two vortices results in the state at $\tilde{t} = 132$, shown in Fig.~\ref{L6a2}b, where the two non-Abelian vortex anyon pairs again annihilate.

\textit{Vortex anyon model---}The cyclic-tetrahedral phase of a spin-2 BEC supports seven distinct fluxon types, labelled as I$_\eta$ - VII$_\eta$ (see Methods). Each of the seven types of fluxon comes with several possible charge labels and taking these into account we obtain the fusion and braiding rules for a complete anyon model. Here we will focus mostly on the fusion of the flux types. The fusion outcomes of the lowest energy fluxons are detailed in the table presented in the Supplementary Figure~S2. Although the type IV$_\eta$ - VII$_\eta$ vortices are non-Abelian anyons, their fusion rules do not preserve the winding number $\eta$ of the anyons, complicating their potential use for topological quantum computation. However, restricting to the set of three fluxons I$_0$, II$_0$ and III$_0$, hereafter referred to as ${\bf 1}, \sigma $, and $\tau$, respectively, results in a concise non-Abelian anyon model. The fusion of two chargeless $\tau$ anyons may result in either a ${\bf 1}$, $\sigma$ or $\tau$ anyon, with the explicit fusion rule $\tau\otimes\tau = N_{\tau\tau}^{\bf 1}{\bf 1} \oplus N_{\tau\tau}^{\sigma}\sigma \oplus N_{\tau\tau}^{\tau}\tau$, where the multipliers $N_{\tau\tau}^{\bf 1}=6$, $N_{\tau\tau}^{\sigma}=6$, and $N_{\tau\tau}^{\tau}=4$ mean that when anyons $a$ and $b$ fuse, they may form a $c$ anyon in $N^c_{ab}$ distinct ways (see Supplemental Material). Note that the 6 distinguishable ways the $\tau$ fluxons can fuse to the flux vacuum are further split by the 4 possible resulting Cheshire charge states and that only one of those 6 fusion channels corresponds to the true vacuum state having both vanishing flux and charge (see Supplemental Material). The non-Abelian $\tau$ anyon is its own antiparticle such that upon fusion, two $\tau$ anyons may annihilate each other. The remaining flux fusion rules of this anyon model are; $\tau\otimes\sigma=\tau$, $\sigma\otimes\sigma={\bf 1}$ and $x\otimes {\bf 1} = x$, where $x \in\{ {\bf 1},\,\sigma,\,\tau\}$. The anyons ${\bf 1}$ and $\sigma$ are Abelian with quantum dimensions $d_{\bf 1}=d_\sigma=1$, respectively. The $\tau$ anyon is the non-Abelian (fluxon) anyon of the theory with a quantum dimension, $d_\tau = 6$, larger than both the Fibonacci and Ising anyon models.

\textit{Topological qubits---}The different fusion outcomes of the anyons define a fusion path, equivalent to a set of topologically distinct states, which can be used for encoding quantum information. We are inspired by the Fibonacci anyon model where the fusion of three anyons allows a topological qubit to be defined as a two-level system plus one non-computational state. In the case of three $\tau$ fluxons, the number of distinct fusion paths in which information could be stored is significantly larger than in the Fibonacci anyon model. Nevertheless, for the sake of demonstration, we consider braiding operations with three fluxons that involve only a subset of the many states in the full fusion space and may therefore be conveniently discussed in terms of effective qubits. A natural choice for the zero state corresponds to the creation of two pairs of $\tau$ fluxons from the true vacuum. The rightmost of the four anyons will not be part of the qubit and will not take part in any braiding processes we consider. Therefore its flux will not be mentioned explicitly in the qubit's state. The zero state of the qubit is then $|0\rangle = \frac{1}{6}\sum_{\gamma_1,\,\gamma_2 \in {\rm III}}|\gamma_1,\,\gamma^{-1}_1,\,\gamma_2\rangle$, corresponding to three $\tau$ anyons with fluxes $\gamma_1,\,\gamma^{-1}_1$ and $\gamma_2$ respectively. A convenient choice for the second qubit state is $|1\rangle = \frac{1}{6}\sum_{\gamma_1,\,\gamma_2 \in {\rm III}}|\gamma_1,\,\gamma_1,\,\gamma_2\rangle$, corresponding to the fusion of the $\tau$ fluxon pair to the $\sigma$ fluxon.

 \begin{figure}[!t]
\centering
\includegraphics[width=0.8\columnwidth]{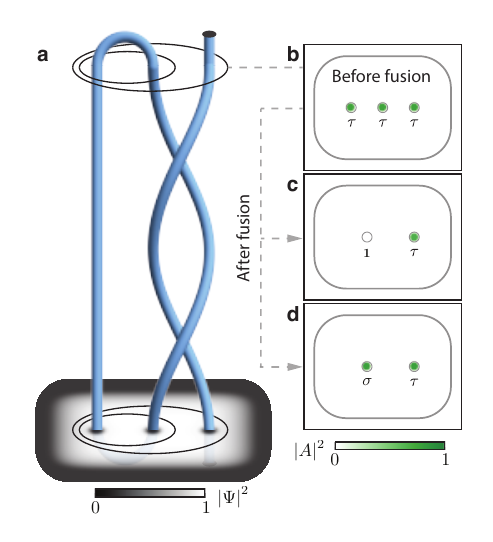}
\caption{\textbf{Single qubit braiding operation}. \textbf{a}, The paths of the three $\tau$ anyons trace out braided world lines enacting a unitary operation on the initial state. Time flows upward. The total condensate density is shown for the initial state. The overlayed concentric ellipses denote the orientation of the qubit as a graphical representation of the bracket notation used in the text. \textbf{b}, Spin-singlet pair amplitude of the qubit just before the fusion. The rounded rectangle marks the boundary of the condensate and the vortex locations are denoted by the circles, the inter-vortex separation is 27$\mu$m. \textbf{c}, a fusion outcome corresponding to the annihilation of the first two anyons as in the $|0\rangle$ state (Supplementary Video 2). \textbf{d}, a fusion outcome corresponding to the non-annihilation of the first two anyons as in the $|1\rangle$ state (Supplementary Video 3). Data in {\bf b}-{\bf d} are thresholded relative to half the maximum value in {\bf{b}} and any maxima within the vortex location markers are mapped to the solid green circles. The raw data is shown in Supplementary Figure~S5. The specific fluxes of the three initial state vortices in (c) are $({\rm III}_{0}^x,-{\rm III}_{0}^x,{\rm III}_{0}^x)$ and in (d) they are $({\rm III}_{0}^x,-{\rm III}_{0}^x,{\rm III}_{0}^y)$.} 
\label{qubit}
\end{figure}

\begin{figure}[t]
\centering
\includegraphics[width=\columnwidth]{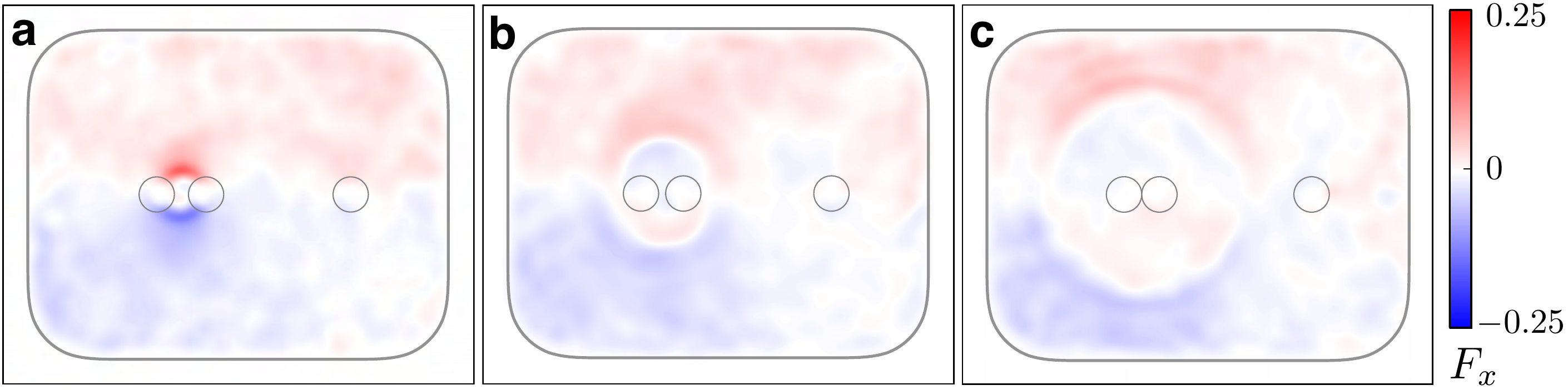}
\caption{\textbf{Signatures of a Cheshire charge}.  Frames (a)-(c) show the $x$-component of the magnetization density of the condensate at the end of the simulation of Fig.~\ref{qubit}(c). The time interval between the frames is $\delta t \approx 16$ms. The circular markers denote the locations of the vortex pinning sites. The expanding ring shaped magnetic soliton structure is emitted due to the fusion of two fluxons, see also Supplementary Video 5.}
\label{charge2}
\end{figure}

Figure~\ref{qubit} demonstrates the action of manipulating the state of such a topological qubit by controllable braiding of the anyons. Initially, the fluxons are prepared in such a way that the first two of them are guaranteed to annihilate upon fusion, as in the $|0\rangle$ state, which in practice could be achieved by nucleating two vortex-antivortex pairs that introduces a fourth, surplus, anyon which is disregarded in this numerical experiment without consequence. 
 
A unitary operation, encoded by the braid in Fig.~\ref{qubit}a, is applied to the fluxons by moving the vortices with the pinning potentials to exchange the second and third anyons within the qubit structure twice. Once the braiding has been completed, a measurement of the state is made by fusing the first and second anyons from the left of the condensate and observing the remaining core structures shortly after the pinning potentials have been withdrawn. Prior to the fusion the three $\tau$ anyons are identified by the green $|A|^2$ cores, as shown in Fig.~\ref{qubit}b. After the braiding, the measurement outcome depends on the topological influence between the exchanged anyons. The braid maps the $|0\rangle$ state to a superposition
\begin{equation}
\sum_{\substack{\gamma_1,\,\gamma_2 \in {\rm III} \\ \gamma_1\gamma_2 =
\gamma_2\,\gamma_1}}
\frac{|\gamma_1,\,\gamma^{-1}_1,\,\gamma_2\rangle}{2\sqrt{3}} \\
 + \sum_{\substack{\gamma_1,\,\gamma_2 \in {\rm III} \\ \gamma_1\gamma_2 \neq
\gamma_2\,\gamma_1}}\frac{|\gamma_1,\,\gamma_1,\,\gamma_2\rangle}{2\sqrt{6}},
\end{equation}
where the two sums contain the combinations of fluxes which braided with trivial and non-trivial topological influence, respectively. The probability $p$ that a measurement would record complete annihilation $p(0)=1/3$ or the formation of a $\sigma$ fluxon $p(1)=2/3$ after the braiding is obtained by projecting the braided superposition state onto the two qubit basis states $|0 \rangle$ and $|1 \rangle$. Prior to the fusion measurement the two possibilities are indistinguishable by any local observation. In general, braiding with respect to this basis would introduce significant leakage into the non-computable fusion paths even for the case of a single qubit. However this is not a real problem as we only restricted to a two dimensional space for illustrative purposes. Any realistic implementation would use the full fusion space for computations.

In the numerical experiments we simulate two specific components of the $|0\rangle$ state, those with fluxes $({\rm III}_{0}^x,-{\rm III}_{0}^x,{\rm III}_{0}^x)$ and $({\rm III}_{0}^x,-{\rm III}_{0}^x,{\rm III}_{0}^y)$, such that the braid acts on these basis states in a deterministic manner. In the first case the exchanged anyons commute so the braid realises a trivial topological influence and the fusion measures the $|0\rangle$ state, shown in Fig.~\ref{qubit}c, characterised by a single green core. However, in the latter case they do not commute so the non-trivial topological influence changes the signs of the anyons and the fusion measures the state $|1\rangle$ of the topological qubit. Such a measurement of the $|1\rangle$ state is illustrated in Fig.~\ref{qubit}d and corresponds to the observation of two green vortex cores, with the additional filled core corresponding to a $\sigma$ anyon formed in the fusion of the two $\tau$ anyons.


\textit{Cheshire charge---}We have discussed the topological qubit at the fluxon level, ignoring the $H$-charges. However, the states considered in the single qubit simulations are $\tau$ flux eigenstates, which correspond to charge superposition states. Here the charge arises as Cheshire charge \cite{Alford1990a,Preskill1990a,Ruostekoski2001a}, which may be revealed when the vortices are annihilated causing the delocalized Cheshire charge to appear. Representations for the four pure (Cheshire) $H$-charge states, less normalisation, that may result from the annihilation of two type $\tau$ fluxons may be expressed in terms of the $\tau$ flux eigenstates as
\begin{align}
|{\rm T0},{\bf 1}\rangle &=\tau_x\tau_{\bar x}+\tau_{\bar x}\tau_x+\tau_y\tau_{\bar y}+\tau_{\bar y}\tau_y+\tau_z\tau_{\bar z}+\tau_{\bar z}\tau_z, \notag\\
|{\rm T1},{\bf 1}\rangle &=\tau_x\tau_{\bar x}+\tau_{\bar x}\tau_x+\theta(\tau_y\tau_{\bar y}+\tau_{\bar y}\tau_y)+\theta^*(\tau_z\tau_{\bar z}+\tau_{\bar z}\tau_z), \notag\\
|{\rm T2},{\bf 1}\rangle &=\tau_x\tau_{\bar x}+\tau_{\bar x}\tau_x+\theta^*(\tau_y\tau_{\bar y}+\tau_{\bar y}\tau_y)+\theta(\tau_z\tau_{\bar z}+\tau_{\bar z}\tau_z), \notag\\
|{\rm T3},{\bf 1}\rangle &=\{(\tau_x\tau_{\bar x}-\tau_{\bar x}\tau_x), ( \tau_y\tau_{\bar y}-\tau_{\bar y}\tau_y), (\tau_z\tau_{\bar z}-\tau_{\bar z}\tau_z)\},
\label{cheshire}
\end{align}
where the $Ti$ refer to the $H$-charges, ${\bf 1}$ is the total flux,  $\theta=e^{i2\pi/3}$ and the notation $\tau_x\tau_{\bar x}$ corresponds to a spinor hosting two vortices, III$^x_0$ and -III$^x_0$. After a Cheshire charge localizes to a $H$-charge, it could reform as a pair of \emph{Alice vortices} or a propagating \emph{Alice ring} \cite{Preskill1990a}. In our single qubit simulations we have observed a propagating ring-shaped soliton structure in the magnetization density of the condensate, Fig.~\ref{charge2}(a-c), see Supplementary Video 5. This observed signature may be related to the phenomenology of Cheshire charges.

We have demonstrated that certain non-Abelian fractional vortices in spinor Bose--Einstein condensates may be non-Abelian anyons and have shown how such vortex anyons could be braided and fused using guiding laser beams. Mochon showed \cite{PhysRevA.69.032306} that anyons based on finite groups that are solvable but not nilpotent are capable of universal quantum computation. Since the binary tetrahedral group does satisfy these criteria, it may be a fruitful platform for developing a universal quantum computer. A method to generate multiple non-Abelian vortices has been outlined in \cite{PhysRevLett.117.275302}. However, to realize such vortices experimentally a series of engineering challenges must be confronted (see Supplemental Material). To ensure the non-Abelian topology, our numerical experiments employ spin interaction strengths which are not currently achievable in experiment. However, a recent proposal by Hurst and Spielman  Ref.~\cite{2018arXiv180908257H} may provide an experimentally realisable pathway for tuning the spin interactions. Promisingly, many additional non-Abelian phases have been predicted for higher spin BEC systems \cite{PhysRevA.85.051606}, which may enable a more accessible experimental route for creating non-Abelian vortex anyons. To surpass the inertial limitations of massive vortices \cite{Simula2018a}, including the adiabaticity speed limit of vortex braiding \cite{Virtanen2001c}, synthetic non-Abelian fluxons could potentially be designed and engineered using novel artificial gauge field techniques \cite{Goldman2014a,Sugawa2018a}. The ability to perform quantum information processing with the non-Abelian vortices is likely compromised by the substantial challenge of creating and maintaining true quantum superpositions with a macroscopic number of atoms in a Bose--Einstein condensate. However, we conclude that non-Abelian vortices in spinor Bose--Einstein condensates hold promise for a tangible demonstration of the underlying principles of topological quantum computation and should be pursued further.

%

\vfill
\textbf{Acknowledgements} This work was supported by the Research Training Program of the Australian Government Department of Education and Training (T.M.), Science Foundation Ireland grants 12/IA/1697 and 16/IA/4524 (J.S.), and the Australian Research Council via Discovery Projects Grant No. DP170104180 and Future Fellowships Grant No. FT180100020 (T.S.). This work was performed in part at Aspen Center for Physics (J.S.,T.S.), which is supported by National Science Foundation grant PHY-1607611. \\
\\
\textbf{Author Contributions} T.S. conceived the idea for the study. T.M. conducted the numerical experiments and wrote the first version of the manuscript. All authors substantially contributed to the development of the theoretical underpinnings and the writing of the manuscript.\\
\\
\textbf{$^*$Corresponding Author} Thomas Mawson (thomas.mawson@monash.edu)\\
\\
\textbf{Competing interests} The authors declare that there are no competing interests.\\
\\
\textbf{Data availability statement} The spinor wave function data that support the findings of this study are available in the figshare repository with the identifier doi:10.26180/5bc087f34bfe0. \\
\\
\textbf{Methods}

\textit{Spin-2 Bose--Einstein condensates---}Within the mean-field theory, a spin-2 Bose--Einstein condensate is described by a vector order parameter $\Psi({\bf r},\,t)$ with five components $\psi_m$, corresponding to the internal spin states of the condensate, indexed by the magnetic quantum number $m=-2,\,-1,\,0,\,1,\,2$ \cite{Kawaguchi2012253}. The Hamiltonian density of the system is 
\begin{equation}
\mathcal{H} = H_0 + \frac{c_0}{2}n(\textbf{r})^2 + \frac{c_1}{2}|\textbf{F}(\textbf{r})|^2 + \frac{c_2}{2}|A(\textbf{r})|^2,
\end{equation}
where the single-particle part, $\sum^2_{m=-2} \psi^*_m[-\hbar^2\nabla^2/2M + \textrm{V}(\textbf{r},\,t)]\psi_m = H_0$, contains the kinetic energy, where $M$ is the atomic mass, and an external potential $\textrm{V}(\textbf{r},\,t)$. Due to the dilute nature of the condensate, the interaction terms can be considered to arise from the s-wave scattering of two-particles in one of the combinations of even total spin $\mathcal{F}$. The scattering can be approximated by a contact interaction with a coupling constant $g_\mathcal{F} = 4\pi\hbar^2a_\mathcal{F}/M$, where $a_\mathcal{F}$ is the s-wave scattering length of the $\mathcal{F}$ spin channel. In the Hamiltonian density the interactions appear as a spin-independent term depending on the total particle density $n(\bf{r})$, and two spin-dependent interactions depending on the spin density vector $\textbf{F}(\textbf{r})=\sum^2_{i,j=-2} \psi^*_i(f_\nu)_{ij}\psi_j$, where $f_\nu$ are the spin-2 Pauli matrices, and the spin-singlet pair amplitude $A(\textbf{r}) = (2\psi_2\psi_{-2}-2\psi_1\psi_{-1}+\psi_0^2)/\sqrt{5}$. The strengths of these interactions are specified by the consolidated coupling constants $c_0= 4\pi\hbar^2(4a_2+3a_4)/7M$, $c_1 = 4\pi\hbar^2(a_4-a_2)/7M$ and $c_2 = 4\pi\hbar^2(7a_0-10a_2+3a_4)/7M$. 

For small external magnetic fields, the $c_1$ and $c_2$ spin interaction strengths determine a phase diagram characterised by the values of $|\textbf{F}(\textbf{r})|$ and $|A(\bf{r})|$. Our focus is on the cyclic-tetrahedral phase which exists for $c_1>0$ and $c_2>0$ and is characterised by $|\textbf{F}(\textbf{r})| =0$ and $|A(\textbf{r})| = 0$. There also exists a ferromagnetic phase for $c_1< 0$ and $c_2>0$, with non-zero $|\textbf{F}(\textbf{r})| $ and $|A(\textbf{r})| = 0$, and an antiferromagnetic phase when $c_1> 0$ and $c_2<0$, with $|\textbf{F}(\textbf{r})| =0$ and non-zero $|A(\textbf{r})|$. Each phase is described by multiple degenerate order parameters connected by  composite gauge and spin rotations. The addition of external magnetic fields lifts this degeneracy due to linear and quadratic Zeeman shifts, further splitting the phase diagram. Spin-2 BECs have thus far been experimentally realised using $^{87}\rm{Rb}$ atoms. Measurements of the scattering lengths of $^{87}\rm{Rb}$ suggest that the condensate lies in the antiferromagnetic phase, though sufficiently close to the phase boundary that uncertainties do not preclude the cyclic-tetrahedral phase \cite{PhysRevA.64.053602,PhysRevLett.92.040402,1367-2630-8-8-152}.

\textit{Non-Abelian vortices---}The types of vortices that can be excited in a BEC are determined by the symmetries of the system's order parameter. The full symmetry group of a spinor Bose gas is $G = \rm{U}(1)\times \rm{SO}(3)$. Each condensate phase corresponds to an order parameter with a distinct discrete broken symmetry described by the order-parameter manifold $\mathcal{M} = G/H$, where $H$ is a finite subgroup of $G$ called the isotropy group whose transformations leave the physical properties of the condensate order parameter invariant. The types of vortices (fluxes) in each ground state phase are classified according to homotopy theory. The classification occurs by mapping each point on a real path encircling a vortex core in the condensate to a path in the corresponding order-parameter manifold. Two paths in the order-parameter manifold are homotopic if they share a base point $x_0\in\mathcal{M}$ and can be smoothly deformed into each other. The homotopic paths, which are the elements of the fundamental group $\pi_1(\mathcal{M},\,x_0)$, form equivalency classes. Each element of $\pi_1(\mathcal{M},\,x_0)$ corresponds to a particular vortex. If $\mathcal{M}=G/H$ is simply connected, and $H$ is a discrete group then $\pi_1(\mathcal{M},\,x_0) \cong H$. Hence, a vortex is equivalently defined as the transformation in $H$ that describes the change in the order parameter along a path enclosing the vortex core. When the fundamental group is non-Abelian, i.e. contains elements that do not commute under the group operation, the corresponding vortices are called non-Abelian vortices. Since $\rm{SO}(3)$ is not simply connected, the vortices are typically characterised by the simply connected covering group $\rm{SU}(2)$. Of the possible subgroups of $\rm{SU(2)}$, the binary dihedral $D^*_n$, binary tetrahedral $T^*$, binary octahedral $O^*$, and binary icosahedral $Y^*$ groups are non-Abelian and correspond to condensate phases with non-Abelian vortices. Such phases occur for spin-$S$ BECs, for $S\geq2$ \cite{PhysRevA.84.053616,PhysRevA.76.013605, PhysRevA.85.051606, PhysRevA.75.023625, PhysRevLett.99.190408, 1751-8121-45-4-045103}. In spin-2 condensates, non-Abelian vortices can be excited in the cyclic-tetrahedral or the biaxial nematic phases, where the latter is realised from the broken degeneracy of the antiferromagnetic phase after the introduction of an external magnetic field.

\textit{Fluxons---}In this work we consider non-Abelian vortex anyon models in the cyclic-tetrahedral phase of the spin-2 BEC with a specific focus on the flux property of the anyons. A representative cyclic-tetrahedral phase order parameter is given by $\Psi = (i,\,0,\,\sqrt{2},\,0,\,i)^{\textrm{T}}/2$ up to a gauge and spin rotation $R = e^{i\phi}e^{-i\theta\boldsymbol{\omega}\cdot\bf{F}}$. The order-parameter manifold of the cyclic-tetrahedral phase is $G/H = \textrm{U}(1)\times\textrm{SU}(2)/T^*$. Hence, the cyclic-tetrahedral phase order parameter is invariant under the 24 elements of the isotropy group $H = T^*$, which consist of discrete composite $\rm{U}(1)$ gauge and $\rm{SU}(2)$ spin rotations. Consequently, each element of $T^*$ corresponds to a distinct vortex. Furthermore, each vortex is categorised into a vortex type (fluxon) according to its equivalency class---the set of fluxes related by the equivalency relation $g \gamma g^{-1}$---where $\gamma$ is the flux of a vortex and $g \in T^*$. The 24 lowest energy vortices, as determined by the $\rm{U}(1)$ phase rotation, are categorised in seven equivalency classes $\rm{I-VII}$. These fluxons are: $(\rm{I})$ the vacuum state; $(\rm{II})$ the integer spin vortex; $(\rm{III})$ the half quantum vortex; $(\rm{IV})$ and $(\rm{V})$ the $1/3$ fractional vortices and $(\rm{VI})$ and $(\rm{VII})$ the $2/3$ fractional vortices. The fluxes of each class are
\begin{align}
(\rm{I})\, &\{(\eta,\,\mathbb{I})\}  \nonumber  \\ 
(\rm{II})\, &\{(\eta,\,-\mathbb{I})\}  \nonumber   \\
(\rm{III})\, &\{(\eta,\,i\sigma_\nu),\,(\eta,\,-i\sigma_\nu)\}  \nonumber   \\
(\rm{IV})\, &\{(\eta+1/3,\,\tilde\sigma), (\eta+1/3,\,-i\sigma_\nu\tilde\sigma)\} \nonumber   \\
(\rm{V})\, &\{(\eta+1/3,\,-\tilde\sigma), (\eta+1/3,\,i\sigma_\nu\tilde\sigma)\}   \nonumber   \\
(\rm{VI})\, &\{(\eta+2/3,\,-\tilde\sigma^2), (\eta+2/3,\,-i\sigma_\nu\tilde\sigma^2)\}  \nonumber   \\
(\rm{VII})\, &\{(\eta+2/3,\,\tilde\sigma^2), (\eta+2/3,\,i\sigma_\nu\tilde\sigma^2)\},  \\ \nonumber
\end{align}
where the $\rm{U}(1)$ component is represented by a winding number $\eta\in\mathbb{Z}$ plus a class specific constant, and the $\rm{SU}(2)$ part by the spin-$1/2$ Pauli matrices $\sigma_\nu$, for $\nu = x,\,y,\,z$, and $\tilde{\sigma}\equiv(\mathbb{I}+i\sigma_x+i\sigma_y+i\sigma_z)/2$. There is no unique way to assign each flux in an equivalency class to each vortex. Each such assignment is related by an isomorphism corresponding to an operation $gXg^{-1}$ of the entire equivalency class by a group element $g\in\rm{T}^*$. As a result, vortices with different fluxes in the same equivalency class are indistinguishable such that the equivalency classes define the indistinguishable fluxon particles---anyons---in our non-Abelian vortex anyon models. 

\textit{Chargeons and dyons---}In addition to the 6 non-trivial fluxons and one vacuum state corresponding to the 7 equivalency classes of $T^*$, there are chargeons that carry a $H$-charge. The algebraic structure of charges and fluxes is determined by the quantum double category theory \cite{deWildPropitius,*deWildPropitius1995a}. The $H$-charges correspond to irreducible representations of the centralizer groups of $T^*$. The centralizers for the fluxon types are $T^*$ (I, II), $Z_4$ (III), and $ Z_6$ (IV - VII) with 7, 4, and 6 irreducible representations, respectively. For a given $U(1)$ winding number, there are 6 non-trivial pure charges (no flux), the same as the 6 pure fluxes. In addition, there are further 29 dyons (charge-flux composites). Thus in total, the cyclic-tetrahedral phase anyon system has one vacuum state and 41 non-trivial particles comprising 6 fluxons, 6 chargeons, and 29 dyons. Additional particles are introduced when the $U(1)$ number is accounted, though many of these particles will behave identically under braiding. The quantum dimension of each anyon is given by product of the order of the associated equivalency class of its flux with the dimension of the irreducible representation of its charge.  

When $H$-charges are braided around fluxes, the result is analogous to the Aharonov-Bohm effect. The flux $\gamma$ will act on the charge $\alpha$ by the matrix $\alpha(\gamma)$ that represents the element $\gamma\in H$, which labels the flux, in the representation $\alpha$ of $H$, which labels the charge. When the group $H$ is commutative, this matrix reduces to a simple phase factor, the usual Aharonov-Bohm phase. A similar result is obtained when braiding dyons, although in this case the acting group is a centralizer subgroup of $H$. Even if no particles with charge are present, charges can still play a role in braiding, since fusion of fluxes may yield anyons with non-trivial charge. For example, two particles (fluxons) each carrying a pure flux may fuse in such a way that the two fluxes annihilate but nevertheless result in a non-trivial $H$-charge. Even if such fluxes are kept apart, they act as carrying a single delocalized charge when a third flux is braided around such a pair of fluxes. The delocalized charge associated with the flux pair is the elusive Cheshire charge, so named after the Cheshire cat, which could disappear, but leave its grin visible \cite{Carroll,Alford1990a,Preskill1990a}. In this work, unless otherwise stated, our focus is on the fluxons of the theory and their fusion and braiding dynamics.  

\textit{Non-Abelian vortices in the biaxial nematic phase---}The biaxial nematic phase also permits non-Abelian vortex anyon models. The representative biaxial nematic order parameter $\Psi = (1,\,0,\,0,\,0,\,1)^{\textrm{T}}/\sqrt{2}$, has an order parameter manifold $G/H = \textrm{U}(1)\times\textrm{SU}(2)/D^*_4$, where $D^*_4$ is the sixteen-element non-Abelian binary dihedral-4 group. Similar to the cyclic-tetrahedral phase, the topological charges are classified into seven equivalency classes. Here the vortex types are: $(\rm{I})$ the vacuum state; $(\rm{II})$ the integer spin vortex; $(\rm{III})$-$(\rm{IV})$ the half quantum vortices; $(\rm{V})$ a half quantum vortex with $\pi/2$ spin rotation; $(\rm{VI})$ a half quantum vortex with $3\pi/2$ spin rotation and $(\rm{VII})$ a half quantum vortex with $\pi$ spin rotation. The corresponding fluxes are
\begin{align}
(\rm{I})\, &\{(\eta,\,\mathbb{I})\}  \nonumber  \\ 
(\rm{II})\, &\{(\eta,\,-\mathbb{I})\}  \nonumber   \\
(\rm{III})\, &\{(\eta,\,\pm i\sigma_x),\,(\eta,\,\pm i\sigma_y)\}  \nonumber   \\
(\rm{IV})\, &\{(\eta,\,i\sigma_z), (\eta,\,-i\sigma_z)\} \nonumber   \\
(\rm{V})\, &\{(\eta+1/2,\,\tilde\sigma), (\eta+1/2,\,-i\sigma_z\tilde\sigma)\}   \nonumber   \\
(\rm{VI})\, &\{(\eta+1/2,\,-\tilde\sigma), (\eta+1/2,\,i\sigma_z\tilde\sigma)\}  \nonumber   \\
(\rm{VII})\, &\{(\eta+1/2,\,\pm i\sigma_x\tilde\sigma), (\eta+1/2,\,\pm i\sigma_y\tilde\sigma)\},  \\ \nonumber
\end{align}
where $\tilde{\sigma} \equiv (\mathbb{I}+i\sigma_z)/\sqrt{2}$. It is noted that the vortices of equivalency classes $\rm{I-IV}$ are the same as those in equivalency classes $\rm{I-III}$ of the cyclic-tetrahedral phase. For the biaxial nematic phase, the $H$-charges correspond to irreducible representations of the centralizer groups of $D^*_4$. The centralizers are $D^*_4$ (I, II), $Z_4$ (III, VII), and $ Z_8$ (IV - VI) with 7, 4, and 8 irreducible representations, respectively. In total, for a given $\eta$, the biaxial nematic phase anyon system has one vacuum state and 45 non-trivial particles comprising 6 fluxons, 6 chargeons, and 33 dyons.

\textit{Non-Abelian vortices in the biaxial nematic phase---}The biaxial nematic phase also permits non-Abelian vortex anyon models. The representative biaxial nematic order parameter $\Psi = (1,\,0,\,0,\,0,\,1)^{\textrm{T}}/\sqrt{2}$, has an order parameter manifold $G/H = \textrm{U}(1)\times\textrm{SU}(2)/D^*_4$, where $D^*_4$ is the sixteen-element non-Abelian binary dihedral-4 group. Similar to the cyclic-tetrahedral phase, the topological charges are classified into seven equivalency classes. Here the vortex types are: $(\rm{I})$ the vacuum state; $(\rm{II})$ the integer spin vortex; $(\rm{III})$-$(\rm{IV})$ the half quantum vortices; $(\rm{V})$ a half quantum vortex with $\pi/2$ spin rotation; $(\rm{VI})$ a half quantum vortex with $3\pi/2$ spin rotation and $(\rm{VII})$ a half quantum vortex with $\pi$ spin rotation. The corresponding fluxes are
\begin{align}
(\rm{I})\, &\{(\eta,\,\mathbb{I})\}  \nonumber  \\ 
(\rm{II})\, &\{(\eta,\,-\mathbb{I})\}  \nonumber   \\
(\rm{III})\, &\{(\eta,\,\pm i\sigma_x),\,(\eta,\,\pm i\sigma_y)\}  \nonumber   \\
(\rm{IV})\, &\{(\eta,\,i\sigma_z), (\eta,\,-i\sigma_z)\} \nonumber   \\
(\rm{V})\, &\{(\eta+1/2,\,\tilde\sigma), (\eta+1/2,\,-i\sigma_z\tilde\sigma)\}   \nonumber   \\
(\rm{VI})\, &\{(\eta+1/2,\,-\tilde\sigma), (\eta+1/2,\,i\sigma_z\tilde\sigma)\}  \nonumber   \\
(\rm{VII})\, &\{(\eta+1/2,\,\pm i\sigma_x\tilde\sigma), (\eta+1/2,\,\pm i\sigma_y\tilde\sigma)\},  \\ \nonumber
\end{align}
where $\tilde{\sigma} \equiv (\mathbb{I}+i\sigma_z)/\sqrt{2}$. It is noted that the vortices of equivalency classes $\rm{I-IV}$ are the same as those in equivalency classes $\rm{I-III}$ of the cyclic-tetrahedral phase. For the biaxial nematic phase, the $H$-charges correspond to irreducible representations of the centralizer groups of $D^*_4$. The centralizers are $D^*_4$ (I, II), $Z_4$ (III, VII), and $ Z_8$ (IV - VI) with 7, 4, and 8 irreducible representations, respectively. In total, for a given $\eta$, the biaxial nematic phase anyon system has one vacuum state and 45 non-trivial particles comprising 6 fluxons, 6 chargeons, and 33 dyons.

\textit{Vortex short-hand notation---}We represent each particular vortex flux with a shorthand notation $\pm \rm{X}^\nu_\eta\equiv(\eta+a_{\rm{X}},\,g^\nu_{\rm{X}})$. The $\rm{X}$ is a Roman numeral corresponding to the class number. The subscript $\eta$ is the winding number of the $\rm{U}(1)$ rotation, which appears in the flux along with a class specific constant $a_{\rm{X}}$. The superscript $\nu$ defines the axis of the $\sigma_\nu$ Pauli matrix generator of the class specific $\rm{SU}(2)$ rotation $g^\nu_{\rm{X}}$, while the sign in $g^\nu_{\rm{X}}$ is determined explicitly by the sign of the label. Example labels for the cyclic-tetrahedral phase vortices are $\textrm{III}^x_0 = (0,\,i\sigma_x)$, $\textrm{IV}_0 = (1/3,\,\tilde{\sigma})$ and $\textrm{-VI}^x_{-1} = (-1/3,\,-i\sigma_x\tilde{\sigma}^2)$. 

\textit{Fusion rules---}The result of fusing two anyons, $a$ and $b$, is given by a fusion rule $a \otimes b = \sum_c N^c_{ab}\,c $, where the sum is over the anyon labels of all possible fusion outcomes. The multiplicity, $N^c_{ab}$, states how many distinguishable ways a particular $c$ anyon can be formed from the fusion of $a$ and $b$ anyons. Similarly, the quantum dimension can be shown to satisfy the relation $d_a d_b=\sum_c N^c_{ab}d_c$. The outcome of fusing two fluxons, $\rm{X}_\eta$ and $\rm{Y}_\nu$, respectively, is determined by a product of their matrix representations following the composition rule $(\pm\textrm{X}^\alpha_\eta)(\pm\textrm{Y}^\beta_\nu)= (\eta+a_{\textrm{X}}+\nu+a_{\textrm{Y}},\,g^\alpha_{\rm{X}} g^\beta_{\textrm{Y}})$.  As a result, their flux-level fusion rules can be determined directly from the fusion tables, which are presented for the cyclic-tetrahedral and biaxial nematic phases in Supplementary Figure ~\ref{S3} and Supplementary Figure ~\ref{S4}, respectively. For example, the fusion of two fluxes of the $\tau$ (III$_0$) anyon can produce a flux of the $\bf{1}$ (I$_0$), $\sigma$ (II$_0$), or $\tau$ (III$_0$) anyon and hence 
\begin{equation}
\tau\otimes\tau=N^1_{\tau\tau}\textbf{1}\oplus N^\sigma_{\tau\tau}\sigma \oplus N^\tau_{\tau\tau}\tau. 
\label{taufuse}
\end{equation}
The  multiplicities are determined as the number of distinct factorizations of the fluxes of each outcome into products of $\tau$ anyon fluxes, equivalent to considering the reverse process where an anyon $c$ splits into anyons $a$ and $b$ in $N^{ab}_c = N^c_{ab}$ ways. As an example, the multiplicity $N^\tau_{\tau\tau} = 4$ since each indistinguishable flux of a $\tau$ anyon can be factorized into 4 distinct products of $\tau$ anyon fluxes. The fusion rules for all the cyclic-tetrahedral and biaxial nematic phase non-Abelian vortex anyons are given at the flux-level in Supplementary Table~\ref{cyclic} and Supplementary Table~\ref{BN}, respectively. The multiplicities are accounted for by the $H$-charges, resulting in richer fusion rules. Considering the $\tau$ fluxon anyon model summarised in Table~{\ref{cyclic}} the charge inclusive fusion rules are:
$\sigma^{(1)}_{\rm T0}\otimes\sigma^{(1)}_{\rm T0}= {\textbf 1}^{(1)}_{\rm T0}$; 
${\textbf 1}^{(1)}_{\rm T0}\otimes{\textbf 1}^{(1)}_{\rm T0}= {\textbf 1}^{(1)}_{\rm T0}$; 
${\textbf 1}^{(1)}_{\rm T0}\otimes\sigma^{(1)}_{\rm T0}= \sigma^{(1)}_{\rm T0}$; 
${\textbf 1}^{(1)}_{\rm T0}\otimes\tau^{(1)}_{\rm Z0}= \tau^{(1)}_{\rm Z0}$;
$\sigma^{(1)}_{\rm T0}\otimes\tau^{(1)}_{\rm Z0}= \tau^{(1)}_{\rm Z0}$;
and
\begin{align}
\tau^{(1)}_{\rm Z0}\otimes\tau^{(1)}_{\rm Z0}=
& {\textbf 1}^{(1)}_{\rm T0} \oplus {\textbf 1}^{(1)}_{\rm T1} \oplus {\textbf 1}^{(1)}_{\rm T2} \oplus {\textbf 1}^{(3)}_{\rm T3} \oplus \notag \\
& \sigma^{(1)}_{\rm T0} \oplus \sigma^{(1)}_{\rm T1} \oplus \sigma^{(1)}_{\rm T2} \oplus \sigma^{(3)}_{\rm T3} \oplus\notag \\
& 2\tau^{(1)}_{\rm Z0} \oplus 2\tau^{(1)}_{\rm Z2} \oplus,
\label{tautau}
\end{align}
where in the above anyon notation $\tau^{(d)}_{\rm c}$ the subscript denotes the $H$-charge of the anyon and the superscript denotes the quantum dimension of the charge. Here $Zi$ and $Ti$ refer to the irreducible representations of the centralizers $Z_4$ and $T^*$, respectively. The particles ${\textbf 1}^{(1)}_{\rm T0}$, $\sigma^{(1)}_{\rm T0}$ and $\tau^{(1)}_{\rm Z0}$ are equivalent to the fluxons I$_0$, II$_0$ and III$_0$, respectively. The physical content of Eq.~(\ref{tautau}) is thus that the fusion of two $\tau$ type fluxons on the left may result in an annihilation to a true vacuum, in either of the two fluxons, one of three chargeons or one of six dyons. It is implicitly understood that charge conservation is maintained in the fusion rules by the presence of Cheshire charge. The Cheshire charge states corresponding to the $Ti$ chargeons in Eq.~(\ref{tautau}) are those in Eq.~(1), which can similarly describe the $\sigma$ dyons after trivially replacing the $\tau$ vortex pair fluxes with those fusing to II$_0$. The Cheshire charges of the $\tau$ dyons are
\begin{align}
|{\rm Z0},\tau\rangle_z &= \{\tau_x\tau_{\bar y}+\tau_{\bar x}\tau_{y}\},\,\{\tau_{y}\tau_x+\tau_{\bar y}\tau_{\bar x}\}, \notag\\
|{\rm Z2},\tau\rangle_z &=\{\tau_x\tau_{\bar y}-\tau_{\bar x}\tau_{y}\},\,\{\tau_{y}\tau_x-\tau_{\bar y}\tau_{\bar x}\},
\end{align}
where $\tau_x\tau_{\bar y}$ corresponds to a spinor hosting two vortices, III$^x_0$ and -III$^y_0$, which fuse to the vortex III$^z_0$.

The result of fusing two chargeons is determined from the tensor product of their representations. The fusion of dyons is more complex as it involves the centralizer groups of the fluxes. Nevertheless, all fusions can be calculated using the tensor product decomposition of the representations of the quantum double $D(H)$ of $H$. This can be done efficiently using the characters of $D(H)$ \cite{0305-4470-32-48-313}.  

In addition to the cyclic-tetrahedral phase $\tau$ anyon model, the biaxial nematic phase additionally supports two concise non-Abelian anyon models. These models consist of the restricted sets of fluxons $\{$I$_0$, II$_0$, IV$_0\}$ and $\{$I$_0$, II$_0$, III$_0$, IV$_0\}$, respectively, for which the fusion rules are given in Supplementary Table~\ref{BN}. In both models, the fluxons I$_0$ and II$_0$ are Abelian anyons with quantum dimensions $d_{\textrm{I}_0}= d_{\textrm{II}_0}=1$, while  III$_0$ and IV$_0$ are non-Abelian anyons with quantum dimensions $d_{\textrm{III}_0}=4$ and $d_{\textrm{IV}_0}=2$, respectively. As with the full fusion rules of the cyclic-tetrahedral anyon models, such as Eq.~(\ref{tautau}), incorporating the $H$-charges will result in a greater number of distinguishable anyons in the model.

\begin{figure*}[h!]
\centering
\includegraphics[width=2\columnwidth]{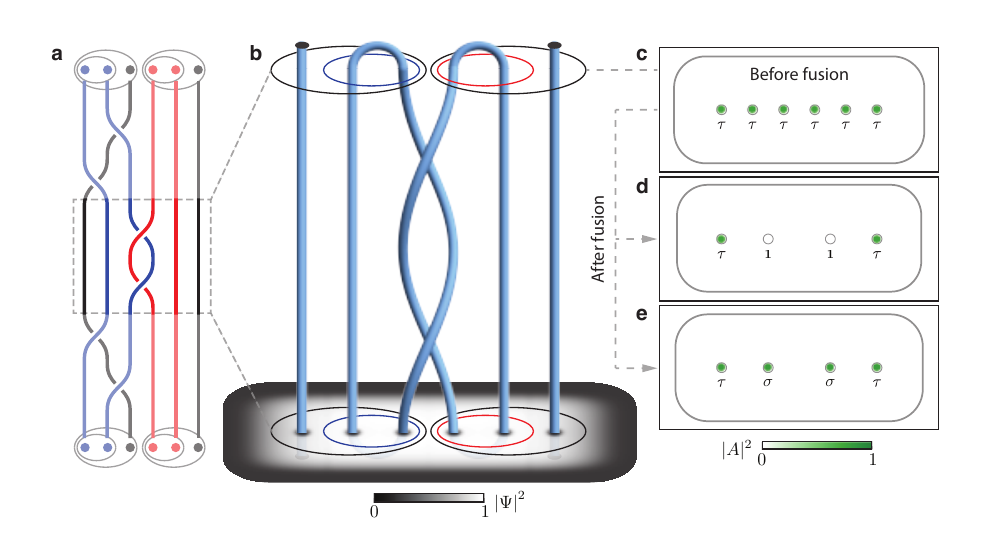}
\caption{\textbf{Two qubit braiding operation}. \textbf{a}, Braid diagram for a unitary operation applied to an initial $|00\rangle$ state where the topologically trivial operations are greyed out. Coloured strands are used to distinguish the measurement anyons of each qubit. \textbf{b}, The paths of the six $\tau$ anyons trace out the topologically non-trivial braid shown in \textbf{a} (Supplementary Video 4). Time flows upward. The total condensate density is shown for the initial state. The overlayed concentric ellipses denote the orientation of the qubits as a graphical representation of the bracket notation used in the text. \textbf{c}, Spin-singlet pair amplitude of the qubits just before the fusion. The rounded square marks the boundary of the condensate and the vortex locations are denoted by the circles. The inter-vortex separation is 27$\mu$m. \textbf{d}, a measurement outcome corresponding to the annihilation of the second and third, and fourth and fifth anyons, counting left to right, as in the $|00\rangle$ state. \textbf{e}, a fusion outcome corresponding to non-annihilation as in the $|11\rangle$ state. Data in {\bf c}-{\bf e} are thresholded relative to half the maximum value in {\bf{c}} and any maxima within the vortex location markers are mapped to the solid green circles. The raw data is shown in Supplementary Figure~\ref{F5s}. The specific fluxes of the six vortices in (d) are $({\rm III}_{0}^z,-{\rm III}_{0}^z,{\rm III}_{0}^y;\,{\rm III}_{0}^z,-{\rm III}_{0}^z,{\rm III}_{0}^y)$ and in (e) they are $({\rm III}_{0}^x,-{\rm III}_{0}^x,{\rm III}_{0}^y;\,{\rm III}_{0}^z,-{\rm III}_{0}^z,{\rm III}_{0}^y)$.}
\label{entglmt}
\end{figure*}

\textit{Entangling topological qubits---}It is straightforward to extend the creation and manipulation of single flux qubits to multiple topological qubits. Figure~\ref{entglmt} demonstrates the action of a unitary braiding operation on two topological qubits comprising six $\tau$ anyons. We consider a $|00\rangle$ initial state, corresponding to both qubits starting in the $|0\rangle$ state. The full braid, shown in Fig.~\ref{entglmt}a, causes an intertwining of the two qubits by tying a topologically non-trivial link in their anyon world lines. The topologically trivial operations in Fig.~\ref{entglmt}a (greyed out) only alter the relative orientation of the two qubits and will have no effect on the fusion outcome. The braid maps the $|00\rangle$ state to a superposition
\begin{align}
& \sum_{\substack{\gamma_{11},\,\gamma_{21},\,\gamma_{12},\,\gamma_{22} \in {\rm III} \\ \gamma_{11}\,\gamma_{12} =
\gamma_{12}\,\gamma_{11}}} 
\frac{|\gamma_{11},\,\gamma^{-1}_{11},\,\gamma_{21};\,\gamma_{12},\,\gamma^{-1}_{12},\,\gamma_{22}\rangle}{12\sqrt{3}} \notag \\
  & + \sum_{\substack{\gamma_{11},\,\gamma_{21},\,\gamma_{12},\,\gamma_{22} \in {\rm III} \\ \gamma_{11}\,\gamma_{12} \neq
\gamma_{12}\,\gamma_{11}}}\frac{|\gamma_{11},\,\gamma_{11},\,\gamma_{21};\,\gamma_{12},\,\gamma_{12},\,\gamma_{22}\rangle}{12\sqrt{6}},
\end{align}
where the two sums contain the combinations of fluxes which braided with trivial and non-trivial topological influence, respectively. The probability $p$ that a measurement would record complete annihilation $p(00)=1/3$ or the formation of two $\sigma$ fluxons $p(11)=2/3$ after the braiding is obtained by projecting the braided superposition state onto the states $|00 \rangle$ and $|11 \rangle$. 

The numerical experiment shown in Fig.~\ref{entglmt}b simulates the action of the topologically non-trivial subsection of the braid on two of the states contributing to $|00\rangle$, those with fluxes $({\rm III}_{0}^z,-{\rm III}_{0}^z,{\rm III}_{0}^y;\,{\rm III}_{0}^z,-{\rm III}_{0}^z,{\rm III}_{0}^y)$ and $({\rm III}_{0}^x,-{\rm III}_{0}^x,{\rm III}_{0}^y;$ $\,{\rm III}_{0}^z,-{\rm III}_{0}^z,{\rm III}_{0}^y)$. The outcomes of measurements corresponding to the first case (d) or second case (e) are shown. Before fusion (c) all anyons have green cores. Upon counting anyons from left to right, when the second and third (in qubit one), and fourth and fifth (in qubit two) anyons are fused, both pairs may either annihilate (d) or leave a green anyon behind (e).  

\textit{Experimental and theoretical feasibility---}Similarly to the case of Ising anyons \cite{Sarma2015a}, the pure fluxons considered here are not capable of universal quantum computation by braiding alone and additional operations involving chargeons will be required to achieve universality \cite{PhysRevA.67.022315,PhysRevA.69.032306}. However, even non-universal anyons may be capable of being used as a quantum memory and even performing certain quantum algorithms with full topological protection, such as a specific Grover search algorithm \cite{LeggettLN} and a calculation of the Jones polynomial at a specific root of unity \cite{Field2018a}. 

Vortices in Bose--Einstein condensates can be manipulated using dynamical pinning potentials generated by focused laser beams \cite{Roberts2014a,Samson2016a}. This enables controlled braiding and fusion of vortex anyons. Condensates containing few hundred vortices can realistically be achieved using current experimental technologies and it may be possible to control them using laser fields morphed with high resolution digital micro-mirror devices (DMDs) \cite{Gauthier:16}. Hence, the creation of 100 high quality topological qubits is a plausible prospect. A potential drawback is the adiabatic speed limit of vortices, in turn limiting the clock speed of such a BEC vortex topological quantum abacus. This issue could perhaps be overcome if robust synthetic non-Abelian fluxons could be created using artificial gauge field techniques \cite{RevModPhys.83.1523,Sugawa2018a}.

The foremost experimental challenge for the realisation of non-Abelian vortex anyons is preparing a Bose--Einstein condensate in a stable non-Abelian ground state phase. Presently, experiments with spinor condensates are limited to phases accessible in the presence of background magnetic field noise and with the natural scattering lengths of available atomic species, since the spin-dependent coupling constant $c_2$ cannot be independently modified using standard Feshbach resonance techniques. Due to uncertainties in the scattering lengths of the rubidium atom, the cyclic-tetrahedral phase may remain an experimentally realisable phase for a spin-2 $^{87}$Rb condensate but likely only in the presence of vanishing external magnetic fields. Moreover, in our numerical experiments we have considered spin interaction strengths stronger than the naturally occuring values for rubidium. This does not affect the topological properties of the anyons but it modifies the spin-healing length (the vortex core size) allowing us to simulate smaller systems. The biaxial nematic phase also hosts non-Abelian vortex anyons and is perhaps a better prospect experimentally, since it may be realised in the presence of an external magnetic field and could potentially be achieved with naturally occurring scattering lengths \cite{PhysRevLett.117.275302}. Furthermore, experimental searches for non-Abelian vortices are not limited to spin-2 condensates with many other non-Abelian phases having been predicted for higher spin BEC systems \cite{PhysRevA.84.053616,PhysRevA.76.013605, PhysRevA.85.051606, PhysRevA.75.023625, PhysRevLett.99.190408, 1751-8121-45-4-045103}. Indeed, the spin-6 condensate, recently realised with erbium atoms \cite{PhysRevLett.108.210401}, has a phase that is symmetric under the non-solvable non-Abelian binary icosahedral group \cite{1751-8121-45-4-045103} and is therefore an attractive prospect for realising universal topological quantum computation employing only fluxons \cite{PhysRevA.67.022315}. 

The most serious issue for the existence of superposition states of the non-Abelian vortex anyons in these systems seems to be the difficulty of creating true quantum superpositions of many atoms. The non-Abelian vortices exist as topological states in the coherent macroscopic spinor wavefunction describing many condensate atoms. Therefore it might be expected that an $N$-atom superposition state realizing a particular fluxon could decohere for example to a state where the atoms realize a definite flux state, instead of a superposition of fluxes. Considering the limit of very small condensate atom numbers the situation may change but also the mean-field description of the bosonic condensates becomes inapplicable.

\textit{Numerical methods---}The numerical experiments are performed by solving the spin-2 Gross-Pitaevski equations \cite{Kawaguchi2012253,PhysRevA.96.033623}, for a condensate of $^{87}\rm{Rb}$ atoms with particle number of either $N=3\times10^5$, for the experiments in Fig.~\ref{L6a2} and ~S4, or $N=1.75\times10^5$, for the single qubit experiment in Fig.~\ref{qubit}, on a spatial mesh of $1024^2$ grid points in 2D. The numerical integration is performed using XMDS2 \cite{dennis2013xmds2} implementing adaptive Runge-Kutta to fourth and fifth order time integration and Fourier spectral methods for spatial integration. The numerical results are presented in non-dimensionalised units $\tilde{t}=1/\omega$, $l=\sqrt{\hbar/2M\omega}$ where $\omega= 2\pi\times5$ Hz and $M$ is the mass of a $^{87}\rm{Rb}$ atom. The dimensionless coupling constant $\tilde{c}_0 = c_0N/\hbar\omega l^2$ is determined from the experimentally measured scattering lengths of $^{87}\rm{Rb}$ \cite{PhysRevA.64.053602,PhysRevLett.92.040402,1367-2630-8-8-152}. Following Kobayashi \textit{et. al} \cite{PhysRevLett.103.115301} the dimensionless spin interaction strengths are chosen as $\tilde{c}_1 = \tilde{c}_2 = 0.5\;\tilde{c}_0$. 

We consider a quasi-uniform condensate in a trapping potential $V_{\rm{ext}}= A\; {\rm{tanh}}[(x/a)^6+(y/b)^6]$, where $A = 50\hbar\omega$, $a = (86,\, 59,\,120)\; \mu$m and $b = (52,\,45,\,52)\;\mu$m, for the numerical experiments in Figs~\ref{L6a2}-\ref{qubit} and Fig~S4, respectively. The pinning potentials used for moving the vortices are modeled as Gaussian laser beams \cite{Simula2005a}
\begin{equation}
V_{\rm{pin}}(r)=\frac{P}{\sqrt{2\pi\sigma^2}}e^{-\frac{(x-x_0)^2+(y-y_0)^2}{2\sigma^2}},
\end{equation}
where $(x_0,\,y_0)$ is the location of the focus point, $P = 63\,\hbar\omega$ and $\sigma=0.5l$. Upon braiding the pinning potentials are moved with an orbital angular frequency $\omega_{\rm{pin}}= \pi^2/2$ rad s$^{-1}$. The vortices are fused by bringing pinning potentials together until they overlap whereupon their amplitudes $P$ are linearly ramped down over a time period of $\approx 127$ms.

\clearpage

\begin{table*}[t]
\caption{Fluxon fusion rules $a\otimes b$ for the non-Abelian vortex anyons in the cyclic-tetrahedral phase. The concise anyon model discussed in the text is highlighted.}
	\centering 
	\begin{tabular}{c|ccccccc} 
    \hline\hline \vspace*{-2.5mm}\\
   & $\rm{I_0}$ & $\rm{II_0}$ & $\rm{III_0}$ & $\rm{IV_0}$ & $\rm{V_0}$ & $\rm{VI_{-1}}$ & $\rm{VII_{-1}}$ \\ \hline \vspace*{-2.5mm} \\
${\bf 1}=\rm{I_0}$ & \cellcolor{Gray}$\rm{I_0}$ & \cellcolor{Gray}$\rm{II_0}$ & \cellcolor{Gray}$\rm{III_0}$ & $\rm{IV_0}$ & $\rm{V_0}$ & $\rm{VI_{-1}}$ & $\rm{VII_{-1}}$\\

 $\sigma=\rm{II_0}$ & \cellcolor{Gray}$\rm{II_0}$ & \cellcolor{Gray}$\rm{I_0}$ & \cellcolor{Gray}$\rm{III_0}$ & $\rm{V_0}$ & $\rm{IV_0}$ & $\rm{VII_{-1}}$ & $\rm{VI_{-1}}$  \\
 
 $\tau=\rm{III_0}$ & \cellcolor{Gray}$\rm{III_0}$ &\cellcolor{Gray}$\rm{III_0}$ & \cellcolor{Gray}$6\rm{I_0}\oplus6\rm{II_0}\oplus4\rm{III_0}$ & $3\rm{IV_0}\oplus3\rm{V_0}$ & $3\rm{IV_0}\oplus3\rm{V_0}$ & $3\rm{VI_{-1}}\oplus3\rm{VII_{-1}}$  & $3\rm{VI_{-1}}\oplus3\rm{VII_{-1}}$ \\ 
 
 $\rm{IV_0}$ & $\rm{IV_0}$ & $\rm{V_0}$ &  $3\rm{IV_0}\oplus3\rm{V_0}$ & $3\rm{VI_0}\oplus\rm{VII_0}$ & $\rm{VI_0}\oplus3\rm{VII_0}$ & $4\rm{I_0}\oplus2\rm{III_0}$ & $4\rm{II_0}\oplus2\rm{III_0}$ \\
 
 $\rm{V_0}$ &  $\rm{V_0}$ & $\rm{IV_0}$ & $3\rm{IV_0}\oplus3\rm{V_0}$  & $\rm{VI_0}\oplus3\rm{VII_0}$  & $3\rm{VI_0}\oplus\rm{VII_0}$  & $4\rm{II_0}\oplus2\rm{III_0}$ & $4\rm{I_0}\oplus2\rm{III_0}$ \\
 
 $\rm{VI_{-1}}$ &  $\rm{VI_{-1}}$ & $\rm{VII_{-1}}$ & $3\rm{VI_{-1}}\oplus3\rm{VII_{-1}}$ & $4\rm{I_0}\oplus2\rm{III_0}$ & $4\rm{II_0}\oplus2\rm{III_0}$ & $3\rm{IV_{-1}}\oplus\rm{V_{-1}}$ & $\rm{IV_{-1}}\oplus3\rm{V_{-1}}$ \\ 
 
 $\rm{VII_{-1}}$ &$\rm{VII_{-1}}$ & $\rm{VI_{-1}}$ & $3\rm{VI_{-1}}\oplus3\rm{VII_{-1}}$ & $4\rm{II_0}\oplus2\rm{III_0}$ & $4\rm{I_0}\oplus2\rm{III_0}$ & $\rm{IV_{-1}}\oplus3\rm{V_{-1}}$ & $3\rm{IV_{-1}}\oplus\rm{V_{-1}}$ \\ \hline\hline
	\end{tabular}
\label{cyclic}
\end{table*}

\begin{figure*}[b]
\centering
\includegraphics[width=2\columnwidth]{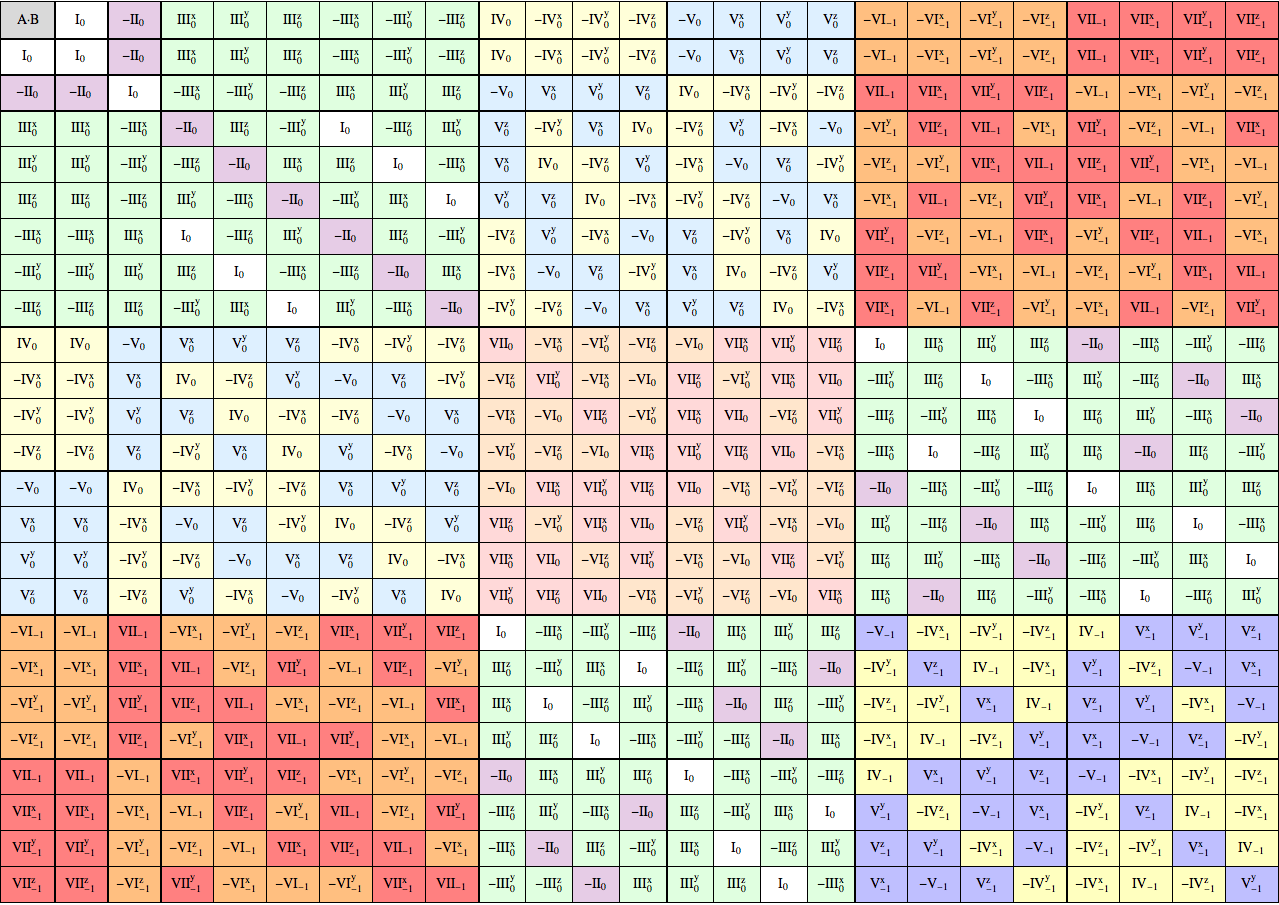}
\caption{\textbf{Fluxon fusion table for the topological charges of the cyclic-tetrahedral phase non-Abelian vortex anyons.} The product A$\cdot$B is ordered with A corresponding to the top row and B to the first column. The thick cell borders divide the regions of each anyon fusion rule. The background colour of each cell signifies the anyon type with the intensity of the shading highlighting the winding number $\eta$.}
\label{S3}
\end{figure*}

\begin{table*}[t]
\caption{Fluxon fusion rules $a\otimes b$ for the non-Abelian vortex anyons in the biaxial nematic phase. The concise anyon models discussed in the text are highlighted.}
	\centering 
	\begin{tabular}{c|ccccccc} 
    \hline\hline \vspace*{-2.5mm}\\
& $\rm{I_0}$ & $\rm{II_0}$ & $\rm{III_0}$ & $\rm{IV_0}$ & $\rm{V_0}$ & $\rm{VI_{0}}$ & $\rm{VII_{0}}$\\ \hline \vspace*{-2.5mm}\\
$\rm{I_0}$ & \cellcolor{LGray}$\rm{I_0}$ & \cellcolor{LGray}$\rm{II_0}$ & \cellcolor{Gray}$\rm{III_0}$ &\cellcolor{LGray} $\rm{IV_0}$ & $\rm{V_0}$ & $\rm{VI_0}$ & $\rm{VII_0}$\\
 
 $\rm{II_0}$ & \cellcolor{LGray}$\rm{II_0}$ & \cellcolor{LGray}$\rm{I_0}$ &\cellcolor{Gray} $\rm{III_0}$ & \cellcolor{LGray}$\rm{IV_0}$ & $\rm{VI_0}$ & $\rm{V_0}$ & $\rm{VII_0}$\\
 
$\rm{III_0}$ & \cellcolor{Gray}$\rm{III_0}$ & \cellcolor{Gray}$\rm{III_0}$ & \cellcolor{Gray}$4\rm{I_0}\oplus4\rm{II_0}\oplus4\rm{IV_0}$ & \cellcolor{Gray} $2\rm{III_0}$ & $2\rm{VII_0}$ & $2\rm{VII_0}$ & $4\rm{V_0}\oplus4\rm{VI_0}$\\ 
 
 $\rm{IV_0}$ & \cellcolor{LGray}$\rm{IV_0}$ & \cellcolor{LGray}$\rm{IV_0}$ & \cellcolor{Gray}$2\rm{III_0}$ & \cellcolor{LGray}$2\rm{I_0}\oplus2\rm{II_0}$ & $\rm{V_0}\oplus\rm{VI_0}$ & $\rm{V_0}\oplus\rm{VI_0}$ & $2\rm{VII_0}$ \\
 
$\rm{V_0}$ & $\rm{V_0}$ & $\rm{VI_0}$ & $2\rm{VII_0}$ & $\rm{V_0}\oplus\rm{VI_0}$ & $2\rm{I_1}\oplus\rm{IV_1}$ & $2\rm{II_1}\oplus\rm{IV_1}$ & $2\rm{III_1}$ \\ 

$\rm{VI_0}$ & $\rm{VI_0}$ & $\rm{V_0}$ & $2\rm{VII_0}$ & $\rm{V_0}\oplus\rm{VI_0}$ & $2\rm{II_1}\oplus\rm{IV_1}$ & $2\rm{I_1}\oplus\rm{IV_1}$ & $2\rm{III_1}$ \\ 

$\rm{VII_0}$ & $\rm{VII_0}$ & $\rm{VII_0}$ & $4\rm{IV_0}\oplus4\rm{V_0}$ & $2\rm{VII_0}$ & $2\rm{III_1}$ & $2\rm{III_1}$ & $4\rm{I_1}\oplus4\rm{II_1}\oplus4\rm{IV_1}$ \\ \hline\hline
	\end{tabular}
	\label{BN}
\end{table*}

\begin{figure*}[b]
\centering
\vspace{30mm}
\includegraphics[width=2\columnwidth]{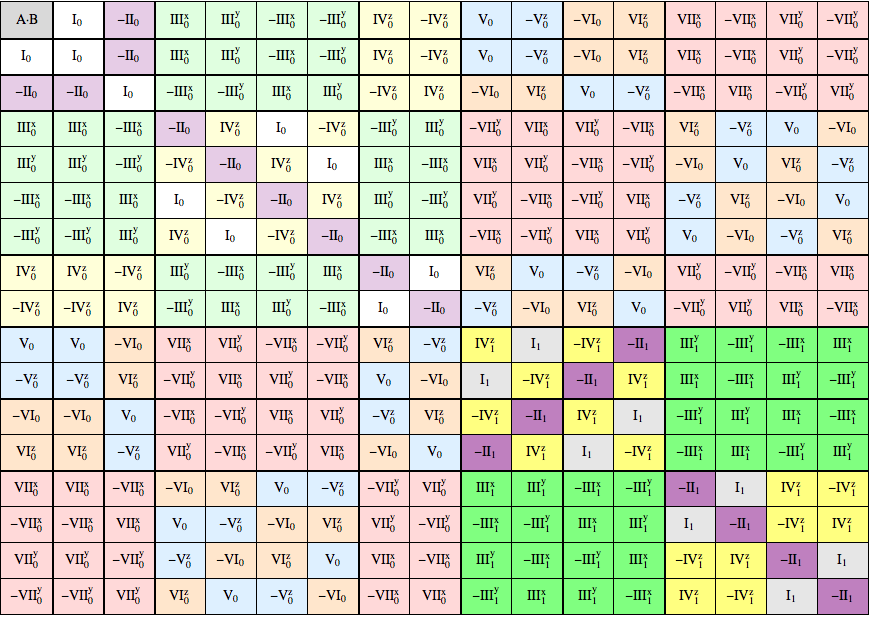}
\caption{\textbf{Fluxon fusion table for the charges of the biaxial nematic phase non-Abelian vortex anyons.} The product A$\cdot$B is ordered with A corresponding to the top row and B to the first column. The thick cell borders divide the regions of each anyon fusion rule. The background colour of each cell signifies the anyon type with the intensity of the shading highlighting the winding number $\eta$.}
\label{S4}
\end{figure*}

\begin{figure}[ht]
\centering
\includegraphics[width=\columnwidth]{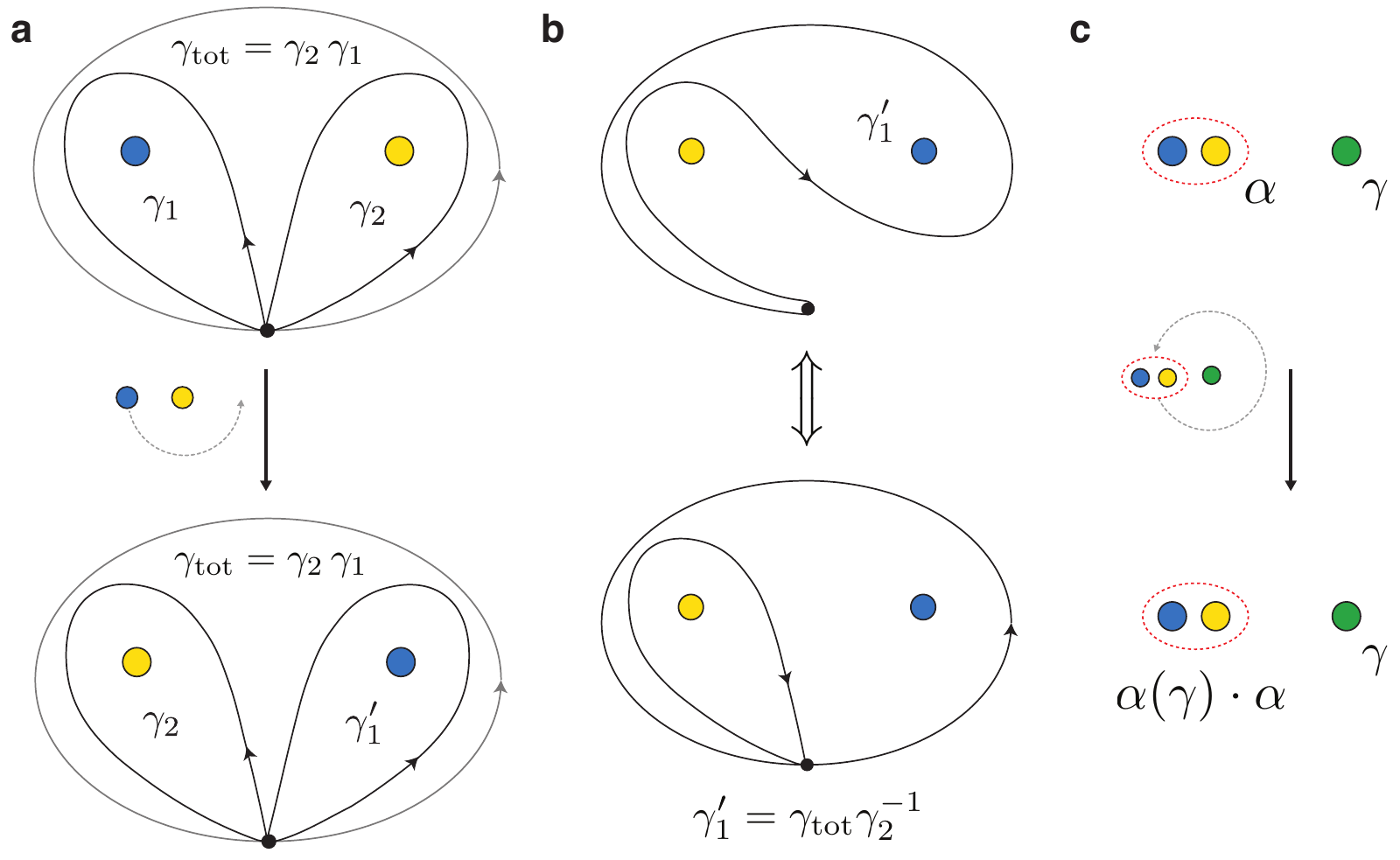}
\caption{\textbf{Topological influence of non-Abelian vortex anyons}. {\textbf a}, Counter-clockwise exchange of two fluxons, denoted by the blue and yellow circles, respectively. Each loop defines a charge. All such loops begin at a base point, denoted by the black circular marker, and encircle either a vortex or vortices in a counter-clockwise sense. A path encircling a vortex clockwise defines the inverse flux. Prior to the exchange the blue and yellow vortices have fluxes $\gamma_1$ and $\gamma_2$, respectively. The loop around both vortices defines the total flux $\gamma_{\rm{tot}} = \gamma_2\gamma_1$. After the exchange the blue fluxon has a new flux $\gamma'_1$. {\textbf b}, The path corresponding to the flux $\gamma'_1$, top row, can be decomposed into the combined paths $\gamma'_1=\gamma_{\rm{tot}}\gamma^{-1}_2$, on the bottom. {\textbf c}, Braid of a fluxon pair with Cheshire charge $\alpha$, denoted by the dotted red ellipse, about a fluxon with flux $\gamma$. The total flux of the fluxon pair commutes with $\gamma$. After the braid the charge acquires a non-Abelian Aharonov-Bohm phase $\alpha(\gamma)$.}
\label{S1}
\end{figure}

\begin{figure}[!h]
\centering
\includegraphics[width=\columnwidth]{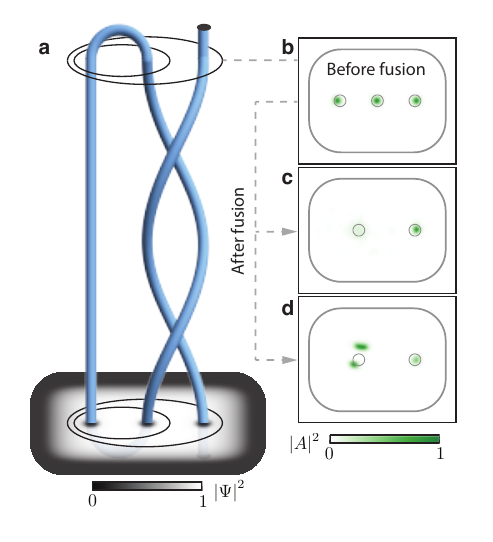}
\caption{\textbf{Single qubit braiding operation}. Raw data for Fig.~2 without thresholding (Supplementary Videos 2 and 3).}
\label{F3s}
\end{figure}
\begin{figure}[!h]
\centering
\includegraphics[width=\columnwidth]{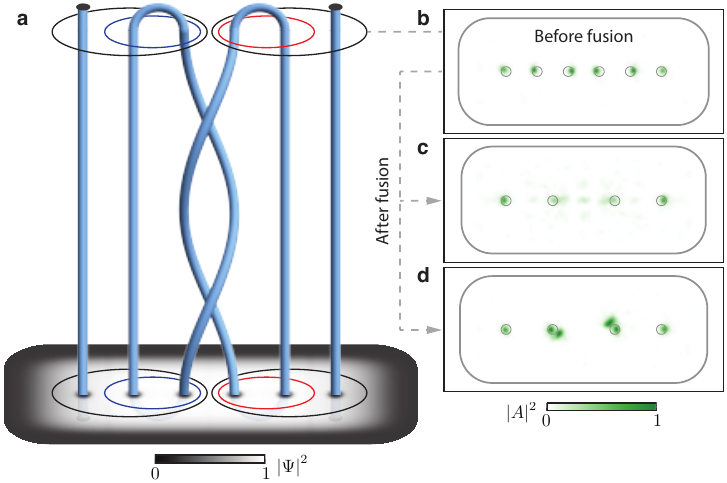}
\caption{\textbf{Two qubit braiding operation}. Raw data for Fig.~\ref{entglmt} without thresholding (Supplementary Video 4).}
\label{F5s}
\end{figure}

\end{document}